\documentclass{iaus}
\usepackage{graphicx}

\title[Planet gaps: Observability with ALMA]
{Planet gaps in the dust layer of 3D proto-\\planetary disks: Observability with ALMA}

\author[J.-F. Gonzalez, C. Pinte, S.T. Maddison \& F. M\'enard]
{J.-F. Gonzalez$^1$, C. Pinte$^2$, S. T. Maddison$^3$ \and F. M\'enard$^2$}

\affiliation{$^1$Universit\'e de Lyon, Lyon, F-69003, France ; Universit\'e Lyon~1, Observatoire de Lyon,\\ 9 avenue Charles Andr\'e, Saint-Genis Laval, F-69230, France ; CNRS, UMR 5574, Centre de Recherche Astrophysique
de Lyon ; \'Ecole Normale Sup\'erieure de Lyon, Lyon, F-69007, France\\
email: {\tt Jean-Francois.Gonzalez@ens-lyon.fr}
\\[\affilskip]
$^2$UJF-Grenoble 1 / CNRS-INSU, Institut de Plan\'etologie et d'Astrophysique de Grenoble, UMR 5274, Grenoble, F-38041, France\\
email: {\tt christophe.pinte@obs.ujf-grenoble.fr, francois.menard@obs.ujf-grenoble.fr}
\\[\affilskip]
$^3$Centre for Astrophysics and Supercomputing, Swinburne University
of Technology,\\ PO Box 218, Hawthorn, VIC 3122, Australia\\
email: {\tt smaddison@swin.edu.au}}

\pubyear{2013}
\volume{299}  
\jname{Exploring the Formation and Evolution of Planetary Systems}
\editors{Brenda Matthews \& James Graham, eds.}
\begin{document}

\maketitle

\begin{abstract}
Among the numerous known extrasolar planets, only a handful have been imaged directly so far, at large orbital radii and in rather evolved systems. The Atacama Large Millimeter/submillimeter Array (ALMA) will have the capacity to observe these wide planetary systems at a younger age, thus bringing a better understanding of the planet formation process. Here we explore the ability of ALMA to detect the gaps carved by planets on wide orbits.
\end{abstract}

\firstsection 
\section{Method}
\label{SectMethod}

In a previous work \cite[(Fouchet \etal\ 2010)]{Fouchet2010}, we ran full 3D, two-fluid Smoothed Particle Hydrodynamics (SPH) simulations of a planet embedded in a gas+dust T~Tauri disk for different planet masses and grain sizes. The gas+dust dynamics, where aerodynamic drag leads to the vertical settling and radial migration of grains, is consistently treated. This produces gaps that are more striking and require a smaller planet mass to form in the dust phase than in the gas. In this work, we pass the resulting dust distributions to the Monte Carlo radiative transfer code \textsf{MCFOST} \cite[(Pinte \etal\ 2006)]{Pinte2006} to construct synthetic images in the ALMA wavebands. We then use the ALMA simulator from the CASA (Common Astronomy Software Applications, \texttt{http://casa.nrao.edu}) package to produce images that include thermal and phase noises for a range of observing parameters. The procedure is detailed in \cite[Gonzalez \etal\ (2012)]{Gonzalez2012}.

\begin{figure}[t]
\begin{center}
\includegraphics[height=5cm]{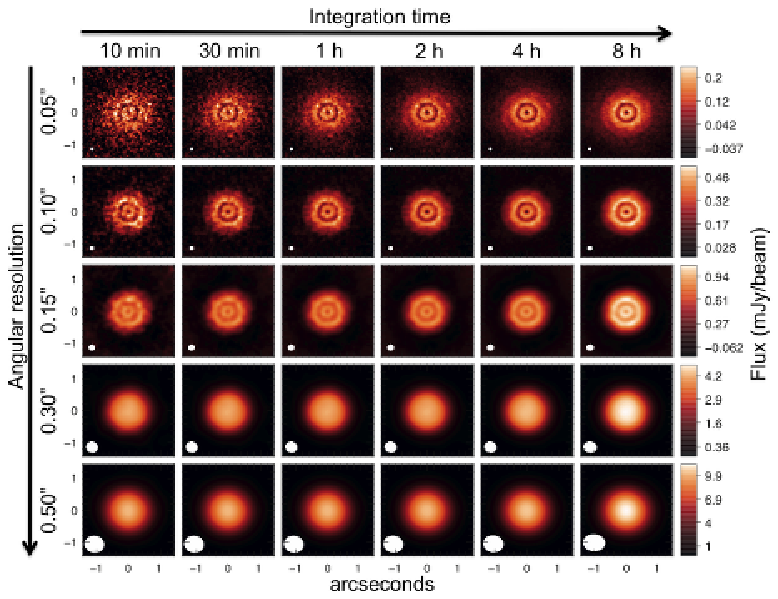}
\hfill
\includegraphics[height=3.8cm]{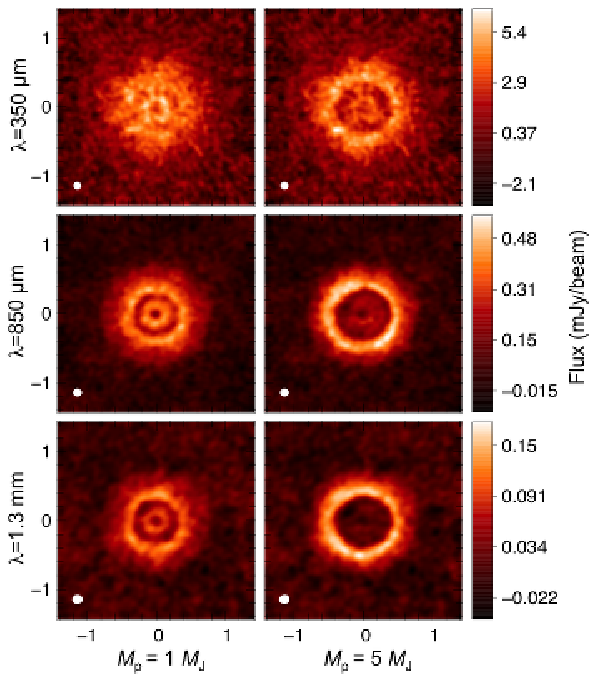}
\hfill
\includegraphics[height=3.8cm]{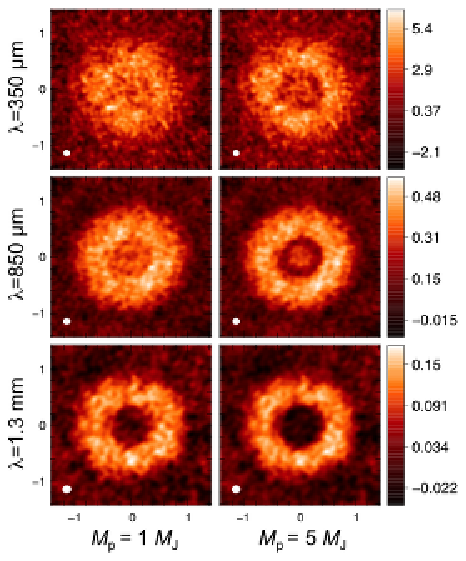}
\caption{Simulated observations of our reference disk at $d=140$~pc, $\delta=-23^\circ$ and $i=18.2^\circ$. Varying integration times and angular resolutions for the 1\,$M_\mathrm{J}$ planet in the dynamic case at $\lambda=850\ \mu$m \textit{(left)}. Optimal parameters ($t=1$~hr, $\theta=0.1''$) for 1 and 5\,$M_\mathrm{J}$ planets in the dynamic \textit{(center)} and well-mixed \textit{(right)} cases.}
\label{FigOpt}
\end{center}
\end{figure}

\section{Results}
\label{SectRes}

We first look among a wide range of parameters for the best compromise between integration time and required S/N and angular resolution allowing to recover the disk features (Fig.~\ref{FigOpt}, left). We find the optimal observing parameters to be $t=1$~h and $\theta=0.1''$ (Fig.~\ref{FigOpt}, center). We also produce images assuming that gas and dust are well mixed (Fig.~\ref{FigOpt}, right) to compare with previous ALMA predictions (e.g.\ \cite[Wolf \etal\ 2002]{Wolf2002}) all made under this hypothesis. This assumption, producing much less defined gaps, with lower contrast, is anyway unrealistic and should not be used.

When including phase noise in median sky quality, our images show a substantial deterioration. However, this should not be an issue when observing in dry weather at short wavelengths and using water vapor radiometers providing real-time corrections.

\begin{figure}[h]
\begin{center}
\resizebox{.6\hsize}{!}{
\includegraphics{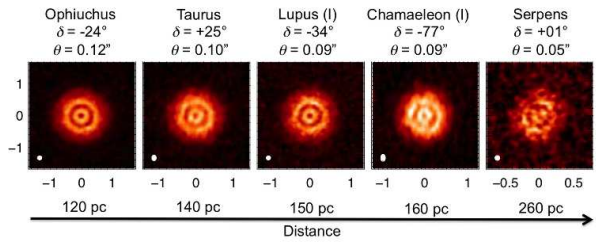}
}
\caption{Simulated observations of disks in star forming regions at different distances and declinations for the 1\,$M_\mathrm{J}$ planet in the dynamic case at $\lambda=850\ \mu$m.}
\label{FigSFR}
\end{center}
\end{figure}

We find that gaps are visible at most disk inclinations: because of dust settling, the disk outer regions are not flared and do not hide them. Planet signatures are fainter in more distant disks but declination has little effect on their detectability thanks to the excellent $uv$ coverage of ALMA (Fig.~\ref{FigSFR}). For more details, see \cite[Gonzalez \etal\ (2012)]{Gonzalez2012}.

\section{Conclusion}
\label{SectConcl}

Gaps will be detectable for lighter planets than anticipated from gas-only simulations and a single 1-hour image can be sufficient to detect one and infer the presence of an unseen planet. Characterizing it requires multi-wavelength follow-up and observing at short wavelength in dry weather is crucial to detect the inner disk and discriminate planet gaps from the inner holes of transition disks. ALMA should routinely observe signposts of planets in nearby star forming regions.

\end{document}